\documentclass[manuscript]{acmart}
\usepackage{array}
\usepackage[center]{subfigure}

\AtBeginDocument{%
  \providecommand\BibTeX{{%
    \normalfont B\kern-0.5em{\scshape i\kern-0.25em b}\kern-0.8em\TeX}}}

\copyrightyear{2021} 
\acmYear{2021} 
\setcopyright{acmlicensed}\acmConference[CHItaly '21]{CHItaly 2021: 14th Biannual Conference of the Italian SIGCHI Chapter}{July 11--13, 2021}{Bolzano, Italy}
\acmBooktitle{CHItaly 2021: 14th Biannual Conference of the Italian SIGCHI Chapter (CHItaly '21), July 11--13, 2021, Bolzano, Italy}
\acmPrice{15.00}
\acmDOI{10.1145/3464385.3464739}
\acmISBN{978-1-4503-8977-8/21/06}

\begin{document}

\title{Reporting Revenge Porn: a Preliminary Expert Analysis}

\author{Antonella De Angeli}
\email{Antonella.DeAngeli@unibz.it}
\affiliation{%
 \institution{Free University of Bozen-Bolzano}
  \streetaddress{Piazza Domenicani 3}
  \city{Bolzano-Bozen}
  \state{}
  \country{Italy}
  \postcode{39100}
}

\author{Mattia Falduti}
\affiliation{%
 \institution{Free University of Bozen-Bolzano}
  \streetaddress{Piazza Domenicani 3}
  \city{Bolzano-Bozen}
  \state{}
  \country{Italy}
  \postcode{39100}
\email{Mattia.Falduti@unibz.it}
}
\author{Maria Menendez Blanco}
\affiliation{%
 \institution{Free University of Bozen-Bolzano}
  \streetaddress{Piazza Domenicani 3}
  \city{Bolzano-Bozen}
  \state{}
  \country{Italy}
  \postcode{39100}
\email{Maria.MenendezBlanco@unibz.it}
}

\author{Sergio Tessaris}
\affiliation{%
 \institution{Free University of Bozen-Bolzano}
  \streetaddress{Piazza Domenicani 3}
  \city{Bolzano-Bozen}
  \state{}
  \country{Italy}
  \postcode{39100}
\email{tessaris@inf.unibz.it}
}

\renewcommand{\shortauthors}{A. De Angeli et al.}

\begin{abstract}
In our research, we focus on the response to the non-consensual distribution of intimate or sexually explicit digital images of adults, also referred as ``revenge porn'', from the point of view of the victims. In this paper, we present a preliminary expert analysis of the process for reporting revenge porn abuses in selected content sharing platforms. Among these, we included social networks, image hosting websites, video hosting platforms, forums, and pornographic sites. We looked at the way to report abuse, concerning both the non-consensual online distribution of private sexual image or video (revenge pornography), as well as the use of ``deepfake'' techniques, where the face of a person can be replaced on original visual content with the aim of portraying the victim in the context of sexual behaviours. This preliminary analysis is directed to understand the current practices and potential issues in the procedures designed by the providers for reporting these abuses.
\end{abstract}




\maketitle

\section{Introduction}

The phenomenon of the non-consensual distribution of intimate or sexually explicit digital images of adults, colloquially known as ``revenge porn'', is still under the spotlight for the toll is taking on members of the online community. Since the phenomenon involves different aspects of personal and social interaction among users of social media, it has been addressed in the literature on several fields of research. 
From the social science perspective, commonalities and differences in motivations accompanying explicit images posted by self-identified straight/gay/lesbian on a popular revenge pornography website are described in \cite{hearn2019}. Moreover, the key three behaviours of image-based sexual abuse (creation, sharing, threats) and the reasons for making and sending self-taken sexual images in young people are pointed out in \cite{powell2019} and in \cite{cooper2016}, respectively. 
From the legal perspective are emerging problems concerning i) the boundaries of the criminal relevance of several behaviours and ii) the effectiveness of the criminal prevention strategies, as reported in \cite{mcglynn2017}. 
%
%
Besides the pure legal discussion, from a technological perspective it emerges the need to identify, remove and conserve (as evidence and for justice reasons) the contested media content reporting revenge pornography, but this result is yet at a preliminary stage \cite{mohanty2019}. Therefore, current weak prevention methods suggest focusing on a strong reaction to this vicious phenomenon. For this reason, it is paramount to analyse the interaction modalities to report the presence of such media content on the platform pages. One of the critical aspects of mitigating the effects of digital abuses is the usability of the procedure to report such abuses, considering also the legal perspective in the technical design. 
Recent research has brought to the fore the paramount need for expertise sharing between HCI and criminal law research \cite{bellini2020}. However, there is a very limited corpus of research in HCI that investigates strategies for implementing legal requirements in the design of digital technologies. Most of the current research focuses on holistic approaches to technology design. For example, a qualitative analysis on the role of digital technologies design in intimate partner violence is presented in \cite{freed2017} and a novel aspect of HCI research about how technologies can interact with justice-oriented service delivery in cases of violence against sex workers is described in \cite{strohmayer2019}. 
This research seeks to contribute to this limited corpus of research by presenting a preliminary analysis of reporting process design from a legal perspective.
%

%
Our approach is elaborated in the context of the project CREEP\footnote{\url{https://creep.projects.unibz.it}} which examines the phenomenon of non-consensual distribution or sharing of intimate or sexually explicit digital images of adults, addressing the difficulties in the law trying to keep up with constantly evolving information technologies, in particular the areas related to social networks. The project considers the interdisciplinary aspects of the problem combining expertises from the criminal law, sociological, and technological fields. Specific attention is given to the Italian context and in particular to the South Tyrolean area, which is a bilingual region (i.e. Italian and German) in the north of Italy. For this reason, the multilingual aspect is taken into account in our analysis.

We aim at investigating how existing technologies and state of the art in internet usage and electronic trust management can be leveraged to mitigate the damages for potential victims, or support victims to report and stop abuses. 
The focus is on the victim perspective, to understand whether best practices can be elaborated and introduced to ease the fight against the phenomenon of revenge pornography. One of the first steps in this direction is to analyse the current practices of content sharing platforms regarding the reporting of abuses. 
In this research, we included also the case of ``deepfake pornography'' as a sexual abuse related to the so-called ``revenge porn'' since it has be leveraged to shame victims by faking their identities on sexual visual contents.

This paper presents a preliminary expert analysis of 45 selected digital platforms describing the state-of-facts of reporting revenge porn cases, also from a legal perspective. 
The paper is organised as follows: in Section~\ref{sec:meth} we describe the methodology of the analysis, with the research question and the adopted procedure. In Section~\ref{sec:res}, we report the first results of the analysis, namely, i) the concrete report modalities, ii) the primary service and the media content present on the platforms, and finally, iii) the jurisdiction of the digital providers. We conclude the paper by discussing the final results in Section~\ref{sec:disc}.
\section{Methodology}\label{sec:meth}
Revenge pornography, as an image-based sexual abuse is usually committed online, and most of the times occurs on digital platforms for content sharing. We aim at describing the state-of-the-facts concerning the reporting of abuses: that is if and how victims can report such abuses, both as users of a platform or as external actors.\footnote{It might be the case that a victim is made aware of the abuse on a platform that (s)he not using.} We elaborate on the following research question:
\begin{quote}
\emph{How complex is the process of reporting image-based sexual abuses from an interaction perspective?}
\end{quote}


 \begin{figure*}[htb]%
       \centering
       \begin{minipage}{.5\textwidth}%
       \centering
         \includegraphics[width=0.5\textwidth]{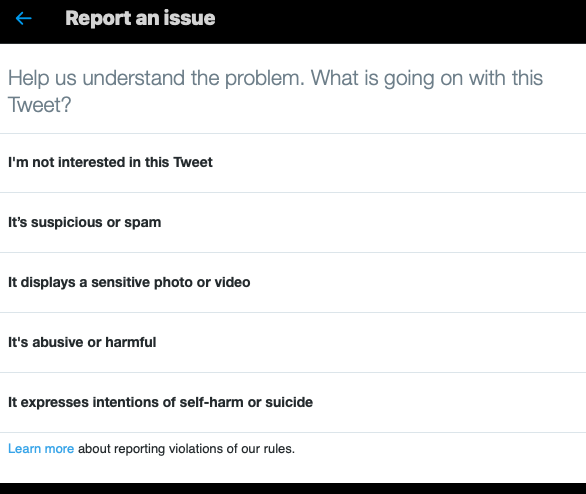}
         \caption{Is it possible to report cases of abuse?}\label{fig:rpt_content}%
       \end{minipage}%
	\begin{minipage}{.5\textwidth}%
	\centering
         \includegraphics[width=0.54\textwidth]{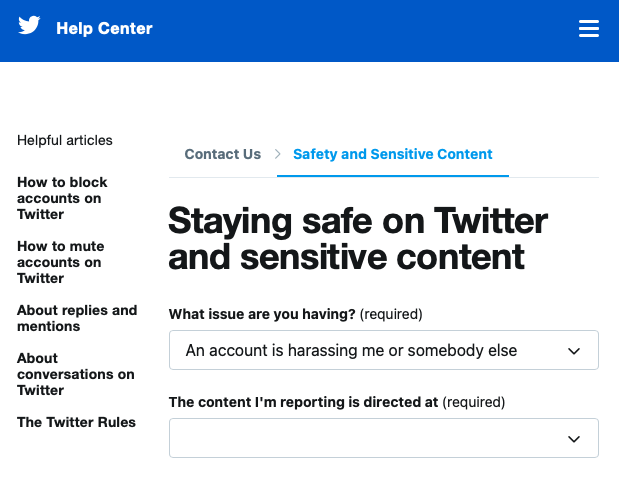}
         \caption{It is possible to report precisely revenge porn cases?}\label{fig:rpt_page}
       \end{minipage}%
     \end{figure*}%

To answer this question we involved a legal expert specialised in criminal law and cybercrime, with a Ph.D.\ in computer science, and co-author of this paper. For each platform, the expert analysed the two modalities to report a case of revenge pornography, namely, i) direct reporting the offending media content (i.e.\ the interaction where the user can select a photo or a video and report it, according to the modalities designed by the platform, as reproduced in Fig.~\ref{fig:rpt_content}), and  ii) indirect reporting the abuse through a general ``report page'' (i.e. a web page where it is possible to report something, for both users or non-users, with a fill-in form, or where platforms display their contacts or information for making a report, as reproduced in Fig.~\ref{fig:rpt_page})

%


\subsection{Procedure}

To the best of our knowledge, there is not a publicly available list of content sharing platforms that can be subject of revenge pornography abuses, therefore we started by collecting a list of platforms to base our analysis on.
The platforms have been selected firstly by including the well known and popular (from the point of view of customer base) ones, and expanding this list by following cross references, posted by users, where the topic of revenge pornography was addressed.
This cross reference following procedure stopped when we reached saturation, meaning that following cross references did not reveal additional relevant information (e.g., the links pointed to landing pages with ambiguous advertisements). 
In total we collected 45 different platforms, starting from the most common social network, image and video hosting websites, and the most popular porn video hosting websites, we identified niche websites, in some cases just dedicated to revenge pornography or deepfake porn content sharing.
We intentionally omit to disclosure the full name of some borderline pornographic websites to avoid indirect publicity of questionable platforms that intentionally permit sharing of revenge pornography. However, in this paper, we mention the most popular pornographic websites, as done in \cite{vallina2019}, and the full list of the analysed platforms is available on request.

For each platform, we performed the following activities.
\begin{itemize}

\item{analysed the primary service, being aware that many platforms host several services, such as messaging, image hosting and video hosting;}
\item{analysed the kind of media hosted w.r.t. \ pornography and the terms of use page;}
\item{searched details concerning the ownerships, nationalities and jurisdiction of the providers;}
\item{verified the possibility to report the media content directly (Fig. \ref{fig:rpt_content}) and if yes, we described the path to the end of the report procedure;}
\item{verified possibility to report the media content indirectly (Fig. \ref{fig:rpt_page}) and if yes, we described the path to the end of the report procedure.}
\end{itemize}

 
\section{Results}\label{sec:res}
In this section, we present the results of our preliminary analysis. Firstly, we describe the analysis of the modalities for revenge pornography. Secondly, we present the analysis on the primary services of the providers, the media hosted w.r.t. pornography and the language of the terms of use page. Finally, we report the details concerning the ownership, nationality and jurisdiction of the providers considered in this preliminary analysis.

\subsection{Interacting with Providers as Crime Answer Actors}
Here we present the results of analysing the two reporting modalities. First, we describe the results of analysing the possibility for
the victim to report directly a specific media content directly (Fig.~\ref{fig:rpt_content}) and the path from the selection of the media content to the end of the report procedure. Secondly, we report whether a dedicated page for indirect reporting of an abuse exists, in which language it is available, if it is reachable from the home page and how is the path from the home page to this page.

\paragraph{Direct Reporting a Specific Media Content.}
We present the actions that a user needs to perform for reporting directly a specific media content, and whether they can identify precisely the criminal nature of the abuse. From our analysis, it emerges that only 24 digital providers permit to report a specific media content (image or video). Differences emerge from the terminology used by the platforms in the procedures, as reported in Table~\ref{tab:path_content}.

\newcolumntype{F}[1]{>{\centering\arraybackslash}m{1,3cm}}

\begin{table*}[htb]
\footnotesize

\renewcommand\arraystretch{1}
 \caption{Direct reporting a specific media content}
    \label{tab:path_content}
\begin{tabular}{p{13cm}|F{}}
\toprule
 {\bf Path with the used terminology}                                                                 			 & {\bf Type}\\
\midrule
Report + Choose (other) + Free text                                                                 						& \\
Report + Free text                                                                                   							 &Free \\
Menu + Report + Choose(1st) + Choose(2nd) + Choose(3rd) + Free text 							&Text\\
Menu + Report + Free text                                                                            						 & \\
\hline
Menu + Report + Choose (Offensive/Copyright/Other)       
& \\                                            			
Menu + Report post + Choose (personal information or fake photo manipulation)                        		& \\
Menu + other + Porn content                                                                          						& \\
Report + Infringes My Rights                                                                         						 &\\
Menu + Report + Report Abuse                                                                         						 &General \\
Menu + Report + Sexual Harassment                                                                    					 &Abuse\\
Report + Pornography/Nudity                                                                         						 &\\
Menu + Report Tweet + It’s abusive or harmful + Includes private information + hacked materials		 &\\
Menu + Report + Sexual Content + Choice (Nudity) + Select Timestamp                                  			 &\\
\hline
Report + Other issues + It’s involuntary pornography + I appear in the image                         			& \\
Report abuse + Intimate content posted without my consent                                           				 &\\
Menu + Report + It’s inappropriate + Nudity or sexual activity + Sharing private images              		 &Revenge \\
Menu + Find support or report photo + Nudity + Sharing private images                                			 &Porn\\
Menu + Embed + Report abuse + Privacy violation + Choose (Yes, my privacy) + My explicit images      	 &\\
Menu + Report + Nudity + Regarding Me                                                                					 &\\
Menu + Report + Nudity, Sexual Activity, Non-Consensual Sexualisation    							&\\
\bottomrule
\end{tabular}
\end{table*}

\begin{figure*}[htb]%
       \centering
       \begin{minipage}{.5\textwidth}%
       \centering
         \includegraphics[width=0.5\textwidth]{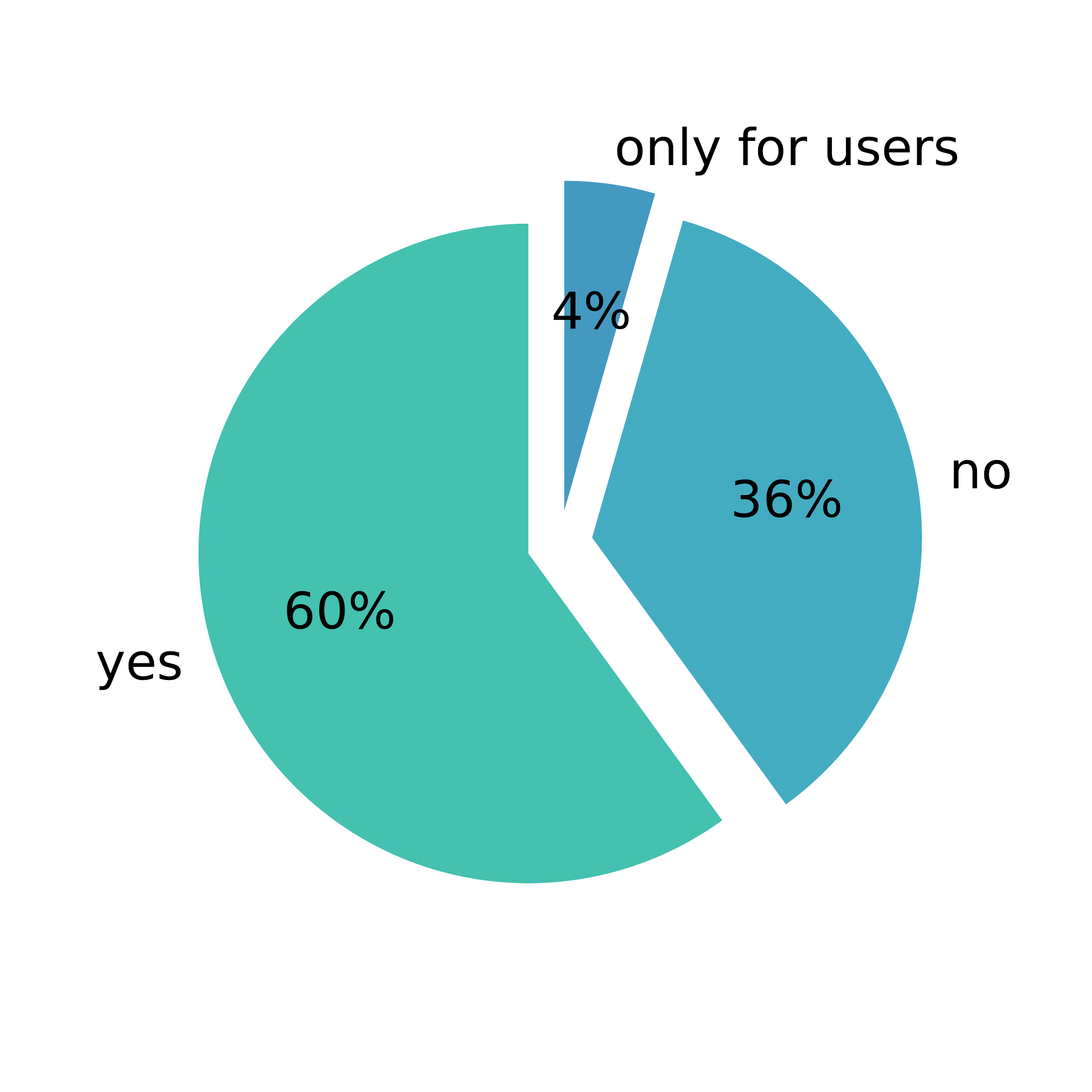}
         \caption{Is it possible to report cases of abuse?}\label{fig:pie1}
       \end{minipage}%
	\begin{minipage}{.5\textwidth}%
	\centering
         \includegraphics[width=0.5\textwidth]{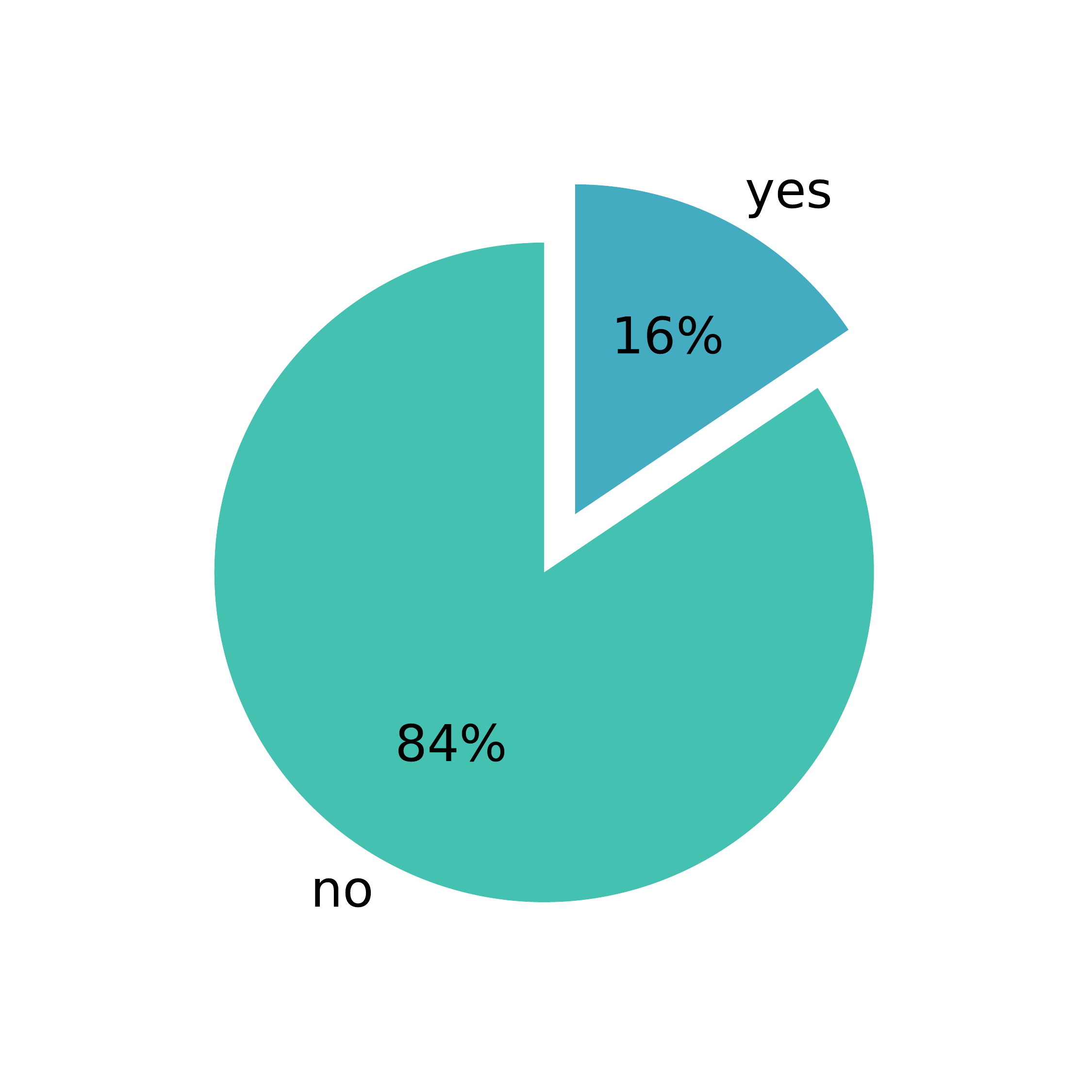}
         \caption{It is possible to report precisely revenge porn cases?}\label{fig:pie2}
       \end{minipage}%
     \end{figure*}

\noindent
Analysing the terminology we identify four types of paths (see Table \ref{tab:path_content}), namely \emph{Free Text} (8) where the victim is facing a text box to fill with the reasons of the report, {\it General Abuse} (9) where the victim can report only a generic abuse, choosing from a list options concerning his/her needs and searching for alternatives or definitions of revenge pornography where maybe revenge pornography could be included somehow, and finally, {\it Revenge Porn} (7), where the victim has access to the dedicated terminology for reporting exactly a revenge porn case. Reporting general pornography or a generic abuse, would permit to report revenge pornography only indirectly. More in detail, considering all the presented results, we could answer if the report of revenge porn cases is possible and how, as depicted in the pie charts reported in Fig.~\ref{fig:pie1} and in Fig.~\ref{fig:pie2} respectively.


\paragraph{Indirect Reporting an Abuse.}
Only a few of the analysed platforms have a dedicated page for reporting cases such as revenge pornography (10). In the other platforms (35), this page is missing at all (10) or is present only an informative page (16), where further instruction for additional actions are described. Other platforms have a page for requesting the removal of (not better specified) content (5), or for reporting a generic inappropriate content (1), or a page for reporting general problems (2). A forum has a page for contacting the moderators (1). 
As said, we analysed also the languages of pages where it is possible to perform an indirect reporting of an abuse and 
only (8) {\tt Twitter, Facebook, Instagram, Snapchat, Dropbox, XVideos, Xnxx and Pornhub} have dedicated report pages available in English, German, and Italian. 
%


\newcolumntype{c}[1]{>{\centering\arraybackslash}m{0,7cm}}

\begin{table*}[htb]
\footnotesize
 \caption{Platforms ranked by steps in reporting revenge pornography}
\renewcommand\arraystretch{1}
\begin{tabular}{p{4cm}|p{9cm}|c{}}
  
\toprule
{\bf Indirect reporting an abuse} & {\bf Direct reporting a specific media content} & {\bf Steps} \\
\midrule
-- & {\tt Twitch} 	& 6\\
\hline
{\tt Instagram} & {\tt Twitter}, {\tt YouTube}, {\tt Tumblr}	& 	5\\
\hline
{\tt Facebook}, {\tt Reddit}	& {\tt Facebook}, {\tt Instagram}, {\tt Snapchat}, {\tt Reddit} 	 & 4\\
\hline
{\tt Snapchat}, {\tt Dropbox}, {\tt XVideos}, {\tt Twitter}  & {\tt XXNX}, {\tt XVideos}, {\tt Viber}, {\tt 4chan}, {\tt Imgur}, {\tt LinkedIn}, {\tt Kik Messenger}, {\tt Line},  {\it omissis} (2) & 3 \\   
\hline
  --	& 	{\tt Flickr}, {\tt TikTok}, {\tt Pornhub}, {\it omissis} (3) & 2 \\  
\hline 
{\tt XXNX},   {\it omissis} (2)	 	& 	-- & 	1	\\   
\bottomrule
\end{tabular}

 \label{tab:path} 
\end{table*}

\subsection{Characteristics of Digital Platforms}
Here we report the details about the primary service of the platforms and about the permission to include pornographic content in the digital platforms pages. Finally, we report the language of the term of use (ToU) documents.
\paragraph{Primary Service.}
The platforms are here classified according to their primary service, even if in some cases, services are overlapping. For instance, we considered a social network the platform that permits the users an interaction based on the sharing of text, images and videos, as its core intent. On the other end, where the primary service was (and still is) identifiable with a particular service, such as video hosting for youtube, we classified the platform according to this. From our analysis, the following primary services emerged: porn video hosting (12), messaging app (8), social network (5), deepfake porn hosting (4), porn forum (3), image hosting (3), forum/image-board (3), video hosting (2), forum (2), deepfake forum (1), microblogging (1), live-streaming hosting (1). 
\paragraph{Pornographic Content.}
With pornographic content, we mean explicit material representing erotic behaviours, intended or not intended to cause sexual excitement, where sexual body parts are not censored. Referring to this definition, we checked if that type of explicit content is permitted or not. From our analysis, it emerged that 34 platforms permit sharing of pornography, whereas 11 platforms do not permit it or we could not find information on their policies on pornographic content after investing a substantial amount of effort. 
\paragraph{Language of the Terms of Use.}
Our research project focuses on a bilingual territory (Italian-German) with the participation of the local community. For these reasons, we report here the languages of the terms of use page, knowing that the large majority of the users ignore it, but understanding as well the relevance of such document from a legal perspective. We counted 25 multilingual web pages, 10 English monolingual, 6 Italian monolingual, 2 bilingual English and Italian, pages containing the terms of use. Two digital platforms do not have a term of use page at all.

\subsection{Providers as Legal Entity}
Platforms are created by providers 
and those exist also off-line, having a nationality and an ownership. Thus, different nationalities imply different legal requirements, jurisdictions and enforcement modalities. Victims of revenge pornography may have the need to contact, interact and establish a connection also formally with the providers, as usually happens when justice authorities or lawyers are dealing with criminal cases committed online. 
As described, best practices for interacting with the providers are missing and victims can react differently against providers. Therefore, this lack of homogeneity in contacting the providers can leave users unsatisfied or even worse, can disorient potential victims when reporting an abuse.

%

\paragraph{Ownership and Nationality.}
Considering the digital providers as legal entity requires an investigation about the nationality, that is imposed by the inevitable relation between the law and its area of application. For this reason, we first record the details concerning the providers ownership and the nationality of those. In some cases, even a seat in a country can influence the applicable law. Many providers do not disclose their ownership for several reasons, and in fact, we were able to determine 
only the following nationalities: Usa 17, Canada 5, Japan 3, Italy 2, Czech 2, China 1 and Russia 1. 
%
%
%


\section{Discussion} \label{sec:disc}
With this preliminary expert analysis, we find differences in the modalities for reporting revenge porn cases on 45 digital providers, in terms of the complexity of the report procedures. The analysis of the page languages, path length and terminology permits us to spot both weak points and unique peculiarities in reporting cases of non-consensual pornography. The expert confirms that the victims of revenge pornography can not rely on common, shared and dedicated best practices to report abuses, elaborated both from a legal and an HCI perspective. Indeed, the analysis confirmed that, even by excluding any mistake in trying to report abuses, i) completing the task of finding a page for reporting indirectly an abuse (if present) can impose up to 5 steps, and ii) completing the task of reporting a specific media content can impose up to 6 steps, and in some cases it is also necessary to write down a narrative of the crime, which can be daunting for a victim of abuse without opportune support.
On this topic, as addressed recently, there is an emerging critical consciousness surrounding the role of HCI in answering crimes. 
With this work, we tried to contribute to the recent discussion about HCI strategies for implementing legal requirements in the design of digital technologies. From our perspective, the design of digital technologies will play an important role in reacting to cyber crimes and an interdisciplinary approach could elaborate successful solutions.


%


\bibliographystyle{ACM-Reference-Format}
\bibliography{main.bib}


\end{document}